\documentclass[a4paper,11pt]{article}
\usepackage{pos}

\usepackage[numbers,sort&compress]{natbib}

\title{Real-time calibrations for future detectors at FAIR}
\author*[a,b]{Valentin Kladov}
\author[a]{Johan Messchendorp}
\author[a,b,c]{James Ritman}

\affiliation[a]{GSI Helmholtzzentrum für Schwerionenforschung GmbH,\\
Planckstraße 1, Darmstadt, Germany}

\affiliation[b]{Fakultät für Physik und Astronomie, Ruhr-Universität Bochum,\\
Universitätsstraße, Bochum, Germany}

\affiliation[c]{Forschungszentrum Jülich GmbH,\\
Wilhelm-Johnen-Straße, Jülich, Germany}

\emailAdd{V.Kladov@gsi.de}
\emailAdd{J.Messchendorp@gsi.de}
\emailAdd{j.ritman@fz-juelich.de}

\abstract{The real-time data processing of the next generation of experiments conducted at FAIR requires a reliable reconstruction of event topologies and, therefore, will depend heavily on in-situ calibration procedures. A neural network-based approach can provide fast real-time calibrations based on continuously available environmental data. We applied this approach to the data obtained from the Drift Chambers of HADES. To enhance regularization we incorporate information about previous environmental states into the Long Short-Term Memory (LSTM) architecture and combine it with Graph Convolutions to account for correlations between different chambers. With the usage of a proposed prediction strategy we achieved stable and accurate predictions, matching the quality of an offline calibration. Moreover, our approach significantly reduces the calibration time, making it well-suited for real-time applications within high-rate data acquisition environments.}

\FullConference{FAIR next generation scientists - 8th Edition Workshop (FAIRness2024)\\
 23-27 September 2024\\
Donji Seget, Croatia\\}

\begin{document}
\maketitle

\section{Introduction}
The Facility for Antiproton and Ion Research (FAIR) plans several advanced experiments, including Compressed Baryonic Matter (CBM) \cite{CBM_description} and antiproton ANnihilation at DArmstadt (PANDA) \cite{PANDA_status_2020}. These experiments will utilize SIS100 acceleration ring and achieve interaction rates significantly higher than at previous GSI experiments, increasing from 20 kHz at HADES to beyond 10 MHz at CBM. To manage the elevated data rates and associated storage costs, CBM will employ online event reconstruction and high-level triggering to select events of interest in real time, aiming to reduce stored data by a factor of 300 \cite{CBM_rates}. The performance of the online event reconstruction will depend strongly on the quality of the calibration parameters, and many of these parameters are sensitive to environmental fluctuations. ALICE is an example of a running experiment in which online calibrations are taking place during synchronous data processing \cite{ALICE_tpcCalib_2017}. \\
For gaseous tracking detectors, online calibrations are particularly critical, and frequent calibrations are required to maintain their performance. An efficient event selection with significant background suppression relies on particle-track reconstruction, which in turn relates to the calibrations of the corresponding systems. The computational complexity of track reconstruction thus creates a problem in achieving adequate real-time precision \cite{Alice_challenges}. \\
One of the promising solutions involves the use of environmental and operational data, such as hit rates and atmospheric pressure, to predict calibration factors without relying on experimental events. Inspired by a Jefferson Lab study \cite{Jefferson}, we develop a neural network-based calibration prediction tool for future detectors at FAIR and test it using data from HADES, a FAIR phase zero experiment \cite{hadesDetector}. Although HADES employs hardware-based triggering and does not use synchronous data processing, it has all the necessary data for the development and proof-of-concept validation of various methods related to online calibrations. \\
This work focuses on the multiwire drift chambers (MDCs) of HADES, a gaseous tracking system comprising 24 spatially separated chambers designed for charged-particle tracking and momentum reconstruction. Operating in the proportional regime, MDCs also enable ionization loss measurements, contingent on the calibrations of the drift velocity and gas amplification factor. These parameters depend on gas properties, high voltage, hit rates, and beam conditions. For the beginning of this study, we target chamber gain calibration, proportional to ionization loss, because it exhibits a strong dependence on environmental conditions. The drift velocity, in contrast, has minimal dependence on the pressure at the working point, which is why its fluctuations remain undetectable with current HADES statistics, leaving it for future investigations.

\section{Method description}
This study is based on data taken with HADES in February 2022 during a beam time with protons ($T=4.5$GeV) on a liquid hydrogen target. In total, around 15,000 runs were taken, each lasting a few minutes. The general idea of a proposed method is to use the available environmental data and to predict the calibration factors for each run in a continuous way using any trainable and automatically re-trainable algorithm. Currently, the method employs a neural network (NN) as a base algorithm and uses the following parameters as input for making a prediction: atmospheric pressure, overpressure, gas concentrations, dew point, high voltage, and trigger rates. These parameters, routinely recorded for monitoring purposes, require minimal adjustments to be incorporated into the online calibration algorithm. \\
In the absence of an official run-by-run calibration for the HADES MDCs, we used a simplified offline algorithm to obtain target values for the network training. For each run and chamber, the time-over-threshold distribution of wire signals is collected without full track reconstruction, then the first noise peak is discarded, and the second peak is fit by a Landau distribution. Although this method is susceptible to noise and secondary signal peaks, it provides sufficient accuracy for preliminary testing. High-quality target calibration, achievable only through complete track reconstruction, was excluded due to computational expense.

\subsection{Neural network architecture}
\begin{figure}[h]
    \centering
    \includegraphics[width=0.95\textwidth]{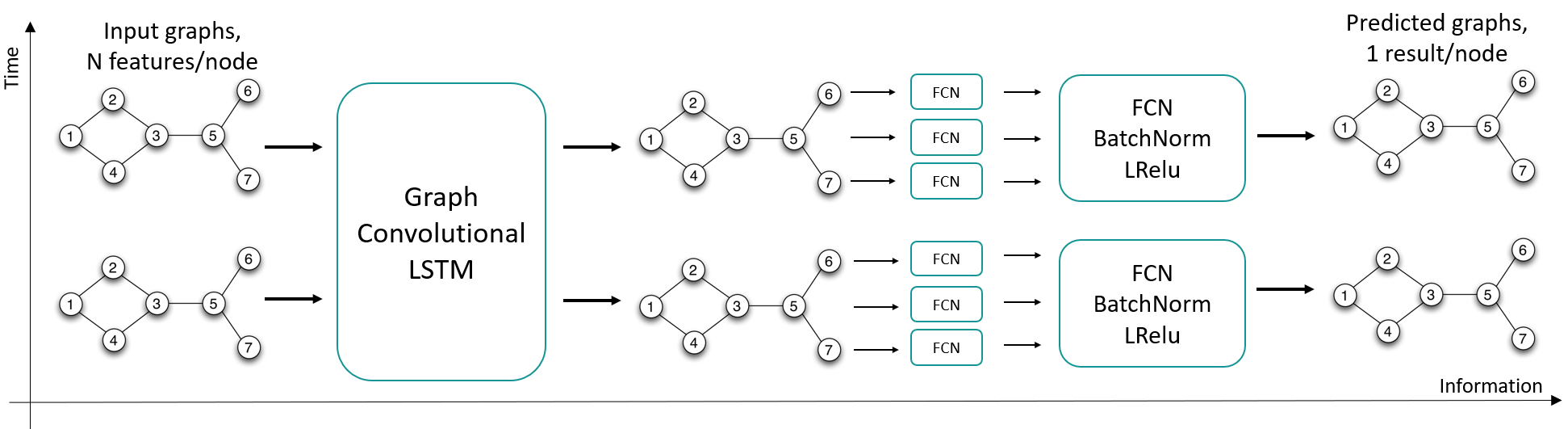}
    \caption{A sketch of the architecture of the network, data flows from bottom to top and from left to right.}
    \label{Fig:architecture}
\end{figure}

Gain fluctuations in the MDC behave differently for different chambers, necessitating individual predictions for each of the 24 chambers. Although some system parameters, such as gas concentrations or atmospheric pressure, are shared between the chambers, they can affect gaseous detectors slightly differently. On the other hand, parameters like overpressure are unique to each chamber, while the general dependence on the absolute value of pressure inside the chamber is similar. To address this, the network was split on a primary Fully Connected Network (FCN) for capturing general dependencies, and an array of small chamber-specific FCN layers. To support future generalization across different detector geometries, the detector was represented as a graph, with chambers acting as nodes. Graph Convolutional Layers (GCNs) then were integrated into the network to capture correlations between adjacent nodes through the links on the graph. Several graph convolution methods were tested that focus on global \cite{graphConvolutionsSpectralFiltering} or local \cite{GCN} correlations. For our case, graph convolutions with spectral filtering performed better than spatial convolutions. \\
Additionally, the introduction of long short-term memory cells (LSTM) into the model was tested, which efficiently handle data sequences \cite{structuredsequencemodelinggraph}. Here the network takes on the input information about several previous environmental states at once and shares information between them. This addition naturally takes into account the smoothness of fluctuations in environment parameters, which usually stay relatively constant during 10-minute time intervals. \\
The resulting architecture of the network that performed the best during testing is presented in Fig.~\ref{Fig:architecture}. GCN and LSTM were combined in one layer by substituting matrix multiplications in the LSTM cells by graph convolutions. The statistics available for the training is limited by the usual beam time at HADES which corresponds to about 10$^{5}$ runs in total. This leads to possible problems with overfitting when the number of parameters in the network is too large. To avoid overfitting, the elements of the network were compressed as much as possible with the exception of the main FCN module. The usage of a common FCN block to predict all 24 target values effectively augments the training data, making the network more robust. 

\subsection{Real-time operation design}
The initial training of the network can be conducted prior to a beam time using data from previous runs or cosmic ray events. After that, the network has to be retrained from time to time with new data to account for gradual detector changes, such as wire degradation. Retraining intervals can be substantially long, on the order of days, judging from the results we observed with HADES. Retraining of the network may take several minutes on a single computing node, but it can be done concurrently with real-time predictions without introducing additional delays. \\
To prepare the method for being applied to future experiments the neural network was integrated into ROOT-based software capable of launching offline calibrations, accessing environmental data, retraining the model, and making calibration predictions automatically as soon as a new run appears on a disk, although the continuous recording of the environmental data allows to make predictions for any time interval. Finally, the method was set up in such a way that the neural network architecture, input parameters, detector geometries or target values can be easily exchanged.

\section{Results}
\begin{figure}[h]
    \centering
    \includegraphics[width=0.95\textwidth]{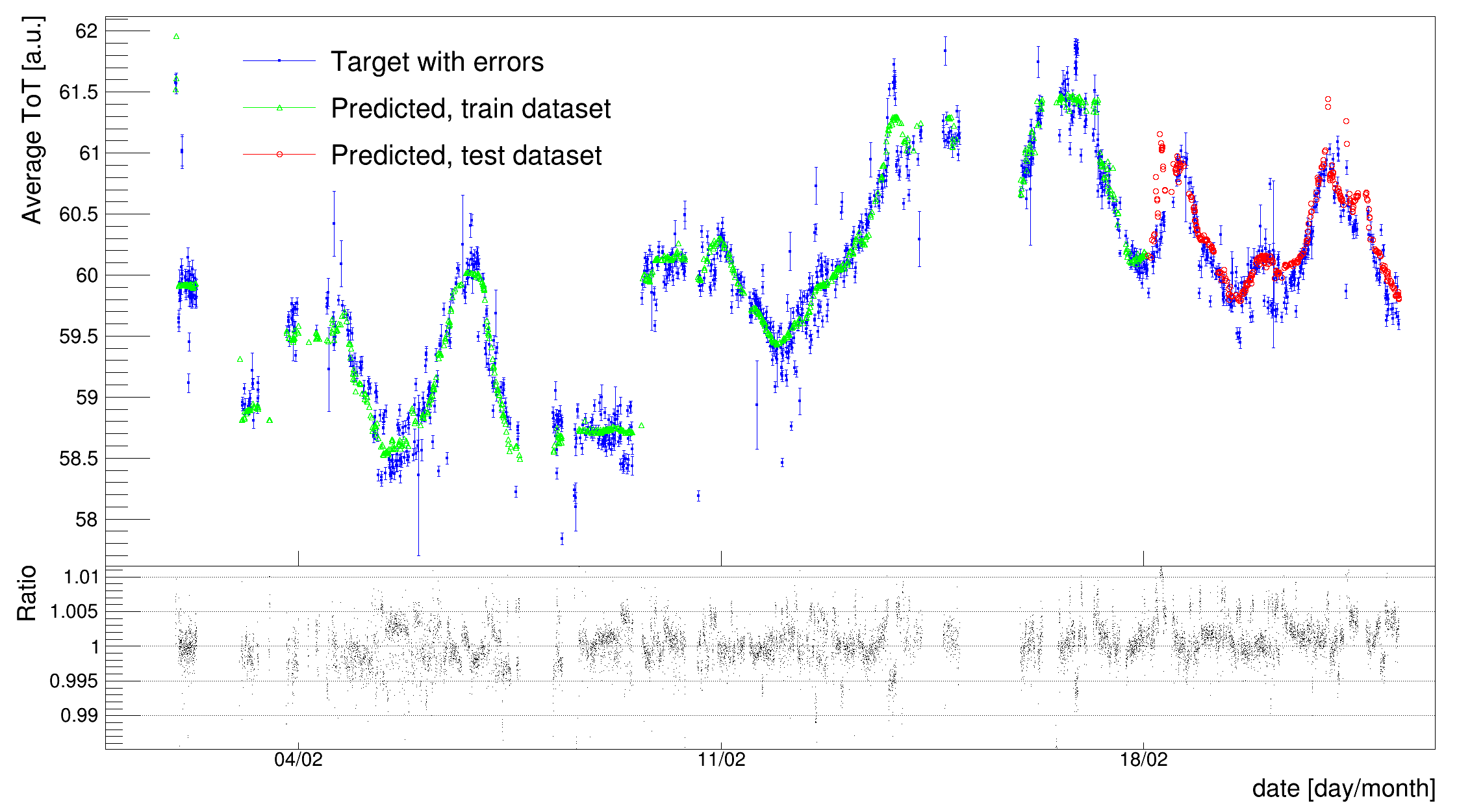}
    \caption{Fluctuations of average measured Time-over-Threshold compared to predicted values and results of initial training for one chamber. The bottom panel shows the ratio of predicted to target values.}
    \label{Fig:performance}
\end{figure}

NN based predictions of calibration coefficients do not require a complete analysis of the event topology and therefore can be done very fast. The average prediction time can be divided into the time needed to read the environmental parameters from the database and the forward pass through the network. The former is on the order of a second, and the network calculation with the currently used architecture takes only 100 ms on a laptop. Additionally, environment parameters have to be measured and stored before making a prediction (around 1 second). These operations are unaffected by track multiplicity or reconstruction complexity and are safely in the time interval where environmental parameters and calibration factors are effectively constant. \\
The quality of predictions was assessed using approximately half of the February 2022 data, with results for one of the chambers shown in Fig.~\ref{Fig:performance}. Time-over-threshold distributions were fitted for each experimental run, with fluctuations in peak positions shown in blue. The network was initially trained on the first portion of the dataset with minimal regularization. It was then retrained with additional regularization and most of the parameters frozen except those of the small FCNs. Finally, predictions were made for the test dataset (red) without further retraining. For clarity, the number of points in the plot was reduced by an order of magnitude. \\
The predicted calibration factors closely follow the offline calibration, confirming the feasibility of the approach. When trained on a sufficiently large dataset, the quality of the network predictions remains stable for several days without retraining, as seen from the figure. No systematic deviations are observed in the test dataset, and no significant overfitting is apparent. Local discrepancies mainly stem from the simplicity of the offline calibration algorithm and a limited number of inputs, restricting the model’s ability to capture higher-order dependencies on the unaccounted parameters.\\

\section{Conclusion and outlook}
A NN-based tool has been developed for the predictions of ionization loss calibration parameters for HADES MDC. We have successfully implemented graph convolutions and recurrent techniques to improve the stability of predictions and achieved accuracy compatible with offline calibrations used for training, thus showing a feasibility of this approach for usage at future FAIR detectors. Possible extensions include an online tuning of the high voltage (HV) which is kept mostly constant during HADES beam times but may be adjusted to maintain stable detector performance and resolution. The neural network can be inverted to identify optimal HV settings that ensure constant calibration factors. This will be studied with cosmic data that will be collected around the next HADES beam time in 2025 with varying MDC HV settings to tune our model.

%\bibliographystyle{apsrev4-1} % For APS-style references
%\bibliographystyle{JHEP} 
%\bibliography{references}

\providecommand{\href}[2]{#2}\begingroup\raggedright\endgroup
  % Use this instead of \bibliography{references}

%\begin{thebibliography}{99}
%\bibitem{article}
%The high-acceptance dielectron spectrometer HADES, Agakichiev, G. and Agodi, C. and Alvarez-Pol, H. and others, The European Physical Journal A, volume 41, 243--277, 2009

%\end{thebibliography}

\end{document}